\documentclass[10pt,conference]{IEEEtran}

\usepackage{cite}
\usepackage{amsmath,amssymb,amsfonts}
\usepackage{algorithmic}
\usepackage{graphicx}
\usepackage{textcomp}
\usepackage{xcolor}
\usepackage{tcolorbox}

\begin{document}
\bstctlcite{IEEEexample:BSTcontrol}

\title{Rule-based hardware-configurable static analysis for quantum programs}

\author{
    Yi-Ting Chen\IEEEauthorrefmark{1}\IEEEauthorrefmark{3},
    Lauren Capelluto\IEEEauthorrefmark{1}\IEEEauthorrefmark{3},
    Ryan Shaffer\IEEEauthorrefmark{1}\IEEEauthorrefmark{3},
    Jeffrey Heckey\IEEEauthorrefmark{2}\IEEEauthorrefmark{3} \\
    \\
    \IEEEauthorblockA{\IEEEauthorrefmark{1}Amazon Braket, New York, USA}
    \IEEEauthorblockA{\IEEEauthorrefmark{2}Amazon Braket, Seattle, USA}
    \IEEEauthorblockA{\IEEEauthorrefmark{3}AWS Quantum Technologies, Seattle, USA}
}

\maketitle
\begin{abstract}
The rapid evolution of quantum hardware necessitates an adaptable static analysis framework for validating quantum programs. In this work, we introduce SHARP, a rule-based static analysis framework designed for OpenQASM that decouples hardware-specific constraints from the validation engine. By employing a rule-based approach, SHARP allows quantum computing services to validate programs against evolving instruction set architectures (ISAs) without having to modify the core analysis engine. SHARP achieves this by ingesting ISA descriptions as standalone data, making the validation engine resilient to hardware updates. Furthermore, the framework employs an analysis-driven validation mechanism that enables multi-stage validation, simplifying rule construction and enhancing extensibility. We demonstrate how SHARP supports hardware-specific validation tasks and discuss its impact on quantum software development, including language decoupling, program-level constraint validation, and feature access control. Our findings suggest that SHARP provides a path toward a scalable and maintainable programming interface for quantum computing.
\end{abstract}


\section{Introduction}
Quantum computing hardware has evolved rapidly. Various architectures are actively being developed, including superconducting qubits\cite{Arute2019,Storz2023,Sete2016,Abdurakhimov2024}, trapped ions\cite{Wright2019,Pino2021,Pogorelov2021}, neutral atoms\cite{Wurtz2023,Henriet2020}, and photonic chips\cite{AbuGhanem2024}, each with unique operational characteristics and challenges\cite{Linke2017,Blinov2021}. Furthermore, some devices expose low-level control of quantum hardware through pulse-programming\cite{Alexander2020,Smith2022} which provides fine-grained control over the qubit state \cite{Kuzmanovi2024,deKeijzer2023}. Additionally, more advanced features such as mid-circuit measurement\cite{Stenger2024}, classical feed-forward control\cite{DeCross2023,Foss-Feig2023}, error detection, and error mitigation are gaining support on more devices.

To innovate quantum applications, it is common for quantum program developers to directly program instructions specific to a particular device, using only the operations natively supported on that device, a type of quantum programming known as verbatim programming or native programming. Taking the native operations as the device's instruction set architecture (ISA), we can draw a comparison to classical computing. Unlike classical computing, where ISAs such as x86, ARM, and RISC-V have matured and stabilized \cite{Waterman2016,Mezger2022,Ali2023}, those for quantum hardware are undergoing rapid evolution. Each new generation may introduce novel instructions or even new data types, modify existing ones, or deprecate outdated operations, making it challenging to maintain a consistent static analysis framework. Furthermore, a quantum computing cloud service like Amazon Braket \cite{braket2019} can host access to diverse hardware. Compatibility issues may arise when software must support multiple ISAs simultaneously \cite{Huang2025}.

\begin{figure}[htbp]
\centering
\includegraphics[width=\linewidth]{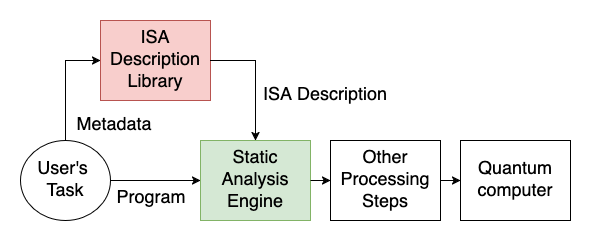}
\caption{Workflow of a service that uses SHARP. When a user submits a quantum task, which includes the quantum program and metadata such as the hardware they want to execute the program on, SHARP finds the ISA description of the hardware and feeds it to the static analysis engine. When adding a new feature, only the ISA description needs to be updated (colored red), while the engine (colored green) does not need to change with each feature addition.}
\label{fig:isa_configurable}
\end{figure}

In this work, we propose a framework that is stable and resilient against the rapid evolution of quantum ISAs. The framework performs the validation step of static analysis. We call this framework \textit{Static Hardware-configurable Analysis with a Rule-based Protocol} or SHARP. Although the idea behind SHARP applies to any program representation, we focus on OpenQASM \cite{Cross2017,Cross2022} in this work. As shown in Fig.~\ref{fig:isa_configurable}, SHARP decouples the ISA description from the static analysis engine. The engine itself has no knowledge of specific hardware; instead, it consumes and processes the hardware description as a standalone piece of data at the time of validation, alongside the program being validated. When a device's existing ISA is modified or new features are supported, no changes are required in the static analysis engine. The onboarding of the feature is done by updating the standalone ISA description. SHARP simplifies the process of onboarding new features, not only reducing the overhead and effort required for implementation, but also making the static analysis layer of the compiler more stable.

The main contributions of this work are as follows:
\begin{itemize}
    \item SHARP is a framework for static analysis of quantum programs that gains resiliency to changes in the quantum ISA by separating the ISA description from the core mechanism of static analysis.
    \item An ISA description is expressed by a rule-based system. The rules are evaluated by an inference engine against program elements when performing static analysis.
    \item The design decouples the language of the quantum program from the core logic of static analysis and the rule-based system in SHARP. An adapter plugin can be used to support new quantum programming languages, without re-implementing the framework.
\end{itemize}

\section{Background}
\subsection{Static analysis against hardware constraints}
Static analysis plays a crucial role in quantum computing services by ensuring program correctness before execution on quantum hardware\cite{Paltenghi2024,Ali2022}. Given the scarcity and limited availability of quantum processors, early detection of program errors through static analysis enhances the user experience by preventing runtime failures\cite{Roberson2010}. Developers receive immediate feedback, reducing debugging time and improving efficiency\cite{Ayewah2007,Wedyan2009}. Moreover, static analysis serves as a safeguard, preventing unreasonable usage that could pose security risks\cite{Charoenwet2024,Zhu2014} or violate hardware constraints\cite{McMahan2019}.

In the context of quantum computing, users can compose an abstract program and use a compiler to convert the program to a hardware ISA. At times, users program specifically to an ISA to have better control over how their programs are run. Therefore, for quantum programs, static analysis must validate not only the syntax and semantics of the input program, but also the hardware-specific constraints. For example,
\begin{verbatim}
rx(angle) qubit;
\end{verbatim}
is an OpenQASM program instruction. The gate, \texttt{rx}, takes an angle as input and applies to a qubit. Given the definition of the valid data types for \texttt{angle} and \texttt{qubit}, validations for \texttt{rx} fit perfectly into grammar-based syntax checkers. However, a hardware ISA may impose additional constraints, such as the acceptable range of angles for each qubit, which is not normally captured by grammars. Therefore, a static analysis framework needs to be hardware-aware, by either hard-coding hardware restrictions or consuming them as input. The latter option makes the framework more flexible.

While quantum compilers like the Qiskit transpiler rely on predefined target data, such as qubit connectivity, they do not handle novel operations beyond its predefined scope \cite{Javadi-Abhari2024,Stefano2024}. Similarly, Q\# provides a ``target profile'' for compiling quantum programs to Quantum Intermediate Representation (QIR) \cite{Svore2018}, but its predefined selection of profiles limits flexibility, preventing developers from specifying arbitrary hardware constraints. For quantum static analysis, the target configurability needs to be generalized beyond a few fixed families of ISAs to accommodate the dynamic and evolving capabilities of quantum hardware.

\subsection{Rule-based software systems}
Rule-based systems have long been recognized as a powerful approach in software design \cite{Hayes-Roth1985,Ligza2006}, particularly for tasks involving decision-making \cite{Miller2019}, validation \cite{Delgado2004}, and expert knowledge representation \cite{Jackson1998,Giarratano2005,Buchanan1984}. The transparency and explainability of rule-based systems make them particularly suitable for applications where clear reasoning and consistent validation against the rules are required \cite{Miller2019,Buchanan1984}.

In the context of program validation, a rule-based approach offers several advantages. By encoding the legal instructions for various ISAs as rules, the system can effectively validate user programs against these predefined criteria \cite{Moller2018}. This design allows for extensibility by simply adding or modifying rules \cite{Jackson1998}, making it suitable for supporting iterations of quantum hardware innovation.

Modern rule-based systems in classical programming further highlight the practicality of this paradigm. One prominent example is ESLint \cite{eslint2013}, a static code analysis tool for JavaScript. ESLint enables developers to define custom linting rules for enforcing code conventions and detecting issues, and its rules can be bundled into configurations and interpreted dynamically. Other systems such as Clang-Tidy for C++ \cite{clang-tidy2025} and Error Prone for Java \cite{errorprone2015} also utilize rule-based mechanisms for static analysis. These tools analyze abstract syntax trees (ASTs) and apply rules that flag violations or potential issues in source code. The success of these rule-based static analyzers in classical domains reinforces the viability of a similar design in quantum programming.

\section{Rule-based static analysis} \label{sec:rule_based_static_analysis}
To address the challenges of constantly evolving hardware ISAs, SHARP performs ISA-aware static analysis with a rule-based system. Rules define constraints on program elements under an ISA, and the rules are targeted by a pass sequencing mechanism. The mechanism performs analysis passes before the main validation passes. Pass sequencing allows for extensible and modular passes. It has been demonstrated to improve performance in code generation \cite{Kahla2023} and code optimization \cite{Grant2000,Abeysinghe2024}. In SHARP, pass sequencing enables multi-stage validation and allows for expressing complex validation logic with simple rules.

In the following, we use an example to introduce the working principles of SHARP. We conceptualize a quantum computer ``QC1'' whose ISA includes a single instruction that is complex enough to demonstrate the expressiveness of SHARP. Real-world hardware that inspires QC1 includes quantum systems from Rigetti and IBM. In this example, we want to validate the OpenQASM snippet below against the ISA of QC1.
\begin{tcolorbox}
\begin{verbatim}
// Example OpenQASM program
OPENQASM 3.0;
valid_gate1 $1;
raw_measure(R1) $1;
\end{verbatim}
\end{tcolorbox}
In this example program, we have one gate \texttt{valid\_gate1} which is already supported by the system. The instruction, \texttt{raw\_measure}, is a novel instruction. We will explore how SHARP performs validation on \texttt{raw\_measure} in two steps:
\begin{enumerate}
    \item An analysis pass traverses the AST of the program and annotates a node if it corresponds to a \texttt{raw\_measure} instruction.
    \item A validation pass enforces the ISA constraints for those nodes annotated as \texttt{raw\_measure} while traversing the AST.
\end{enumerate}
In the subsections below, we explain how to construct the constraints from the hardware as rules and then we explain how to sequence the rules and their corresponding actions to validate the \texttt{raw\_measure} instruction.

\subsection{Hardware constraints as SHARP rules} \label{subsec:ISA_constraints_as_rules}
The raw measurement instruction returns the raw in-phase/quadrature (I/Q) readout values \cite{Heinsoo2018} on a qubit instead of the binary readout classification result of \texttt{0} or \texttt{1}. Assume QC1 has the following specification for the \texttt{raw\_measure} instruction:
\begin{enumerate}
    \item Constraint on qubits: It takes a qubit identifier as input. Only qubits \texttt{\$1}, \texttt{\$2}, and \texttt{\$3} on QC1 support this instruction.
    \item Constraint on readout resonators: It takes a resonator identifier as input. The two resonators on the device are labeled \texttt{R1} and \texttt{R2}. While \texttt{\$1} physically connects to \texttt{R1} and \texttt{\$2} to \texttt{R2}, \texttt{\$3} connects to both \texttt{R1} and \texttt{R2}.
    \item Return value type: It returns two real numbers for the I/Q readout values.
\end{enumerate}

We will create a rule to represent this instruction specification. First, we define the \texttt{Rule} object, which is constructed with ``attribute'' and ``kernel'' arguments. We use the following pseudo-code syntax to express a \texttt{Rule} object. The same pseudo-code syntax is used throughout this paper.
\begin{verbatim}
rule = Rule(
    attribute=attribute,
    kernel=kernel
)
\end{verbatim}
The ``attribute'' argument is an abstract object that defines the properties of a program element, such as the value of an input to a gate, to check against. The ``kernel'' argument defines how to check the attribute. For example, it may check whether an attribute is equal to a target value or whether each element of the attribute has specific types.

More complex constraints can be constructed by stacking \texttt{Rule} objects. Common ways to stack rules include using \texttt{ALL}, which returns true if all rules are true; \texttt{ANY}, which returns true if one of the rules is true; and \texttt{CONDITIONAL}, which evaluates the condition and then the rules if the condition is evaluated to be true. For brevity, we will refer to a \texttt{Rule} object and a stack of rules both as a ``rule.'' Below, we construct a rule for each of the three constraints.

\subsubsection{Constraint on qubits}
The constraint on the qubits can be expressed as the following rule:
{\small
\begin{verbatim}
qubit_val_rule = Rule(
    attribute=Instruction.qubits,
    kernel=Kernel.in([$1, $2, $3])
)
\end{verbatim}
}

\subsubsection{Constraint on readout resonators}
The constraint for the resonator is more complex, as it is a constraint between the two input arguments. It can be expressed as follows:
{\small
\begin{verbatim}
resonator_val_rule = ANY(
    ALL(
        Rule(
            attribute=Instruction.qubits,
            kernel=Kernel.equal($1)
        ),
        Rule(
            attribute=Instruction.inputs,
            kernel=Kernel.equal(R1)
        )
    ),
    ALL(
        Rule(
            attribute=Instruction.qubits,
            kernel=Kernel.equal($2)
        ),
        Rule(
            attribute=Instruction.inputs,
            kernel=Kernel.equal(R2)
        )
    ),
    ALL(
        Rule(
            attribute=Instruction.qubits,
            kernel=Kernel.equal($3)
        ),
        Rule(
            attribute=Instruction.inputs,
            kernel=Kernel.in([R1, R2])
        )
    ),
)
\end{verbatim}
}

\subsubsection{Return value type}
The hardware specification tells us what return type to expect. We can also represent this as a constraint to assert that developers are handling the return type correctly:
{\small
\begin{verbatim}
return_val_rule = ALL(
    Rule(
        attribute=Instruction.returns.length,
        kernel=Kernel.equal(2)
    ),
    Rule(
        attribute=Instruction.returns.type,
        kernel=Kernel.all_equal(FloatType)
    )
)
\end{verbatim}
}

The overall specification for the \texttt{raw\_measure} feature can be constructed by stacking the three rules that we have created, denoted as \texttt{raw\_measure\_rule} below.
{\small
\begin{verbatim}
raw_measure_rule = ALL(
    qubit_val_rule,
    return_val_rule,
    resonator_val_rule
)
\end{verbatim}
}

\subsection{Analysis-driven validation} \label{subsec:Analysis_driven_validation}
Now we have a rule that can validate the \texttt{raw\_measure} instruction. However, it is not the only rule for the program. Let's assume each instruction of QC1, e.g., \texttt{valid\_gate1}, is subject to a rule. These rules should conditionally apply to specific types of instructions. The conditions for applying validation rules are also defined as rules. We call these conditions ``analysis rules'' to distinguish them from validation rules. Validation rules and analysis rules trigger different actions: a validation rule throws an error if the rule fails, while an analysis rule produces information about whether to apply a validation. A \texttt{Pass} is a construct used to encapsulate the rule-action pair, where the action only takes effect if the rules in the pass are evaluated to True.

{\small
\begin{verbatim}
    pass = Pass(
        rule = rule,
        action = action,
    )
\end{verbatim}
}
In order to chain an analysis pass to a validation pass, the result of the analysis is stored in the AST nodes. The action of an analysis pass reads from a node, performs analysis using its rules, and then caches the analysis results back to the node. The information in the node determines the behavior of subsequent validation or analysis passes. On the other hand, a validation pass checks its rule against an AST node and throws an error if the node represents an invalid program element. A node contains a collection of key-value pairs, taking the form of:
{\small
\begin{verbatim}
ast_node_data = {
    key1: value1,
    key2: value2,
    ...
    keyK: valueK
}
\end{verbatim}
}
Below, we construct the analysis and validation passes to validate \texttt{raw\_measure} instructions.

\subsubsection{Define the analysis rule}
The validation rule we constructed in Section~\ref{subsec:ISA_constraints_as_rules} should only target the \texttt{raw\_measure} instructions. To construct the analysis pass that implements rule targeting, we first create the analysis rule which evaluates to True only for AST nodes that represent \texttt{raw\_measure} instructions.
{\small
\begin{verbatim}
analysis_rule = ALL(
    Rule(
        attribute=Instruction.name,
        kernel=Kernel.equal("raw_measure")
    )
)
\end{verbatim}
}

\subsubsection{Define the analysis pass}
The analysis pass action sets the \texttt{is\_raw\_measure} field in each AST node to True if the node satisfies the analysis rule.
{\small
\begin{verbatim}
analysis_pass = Pass(
    rule=analysis_rule,
    action=set(
        ast_node_data,
        key="is_raw_measure",
        value=True
    )
)
\end{verbatim}
}

\subsubsection{Define the validation pass}
We construct the validation pass conditioned on the truth values of \texttt{is\_raw\_measure} and \texttt{raw\_measure\_rule}. The first check is expressed as a rule:
{\small
\begin{verbatim}
condition_rule = Rule(
    attribute=ast_node_data.get("is_raw_measure"),
    kernel=Kernel.equal(True)
)
\end{verbatim}
}
Together, the validation pass is
{\small
\begin{verbatim}
validation_pass = Pass(
    rule=CONDITIONAL(
        condition=condition_rule,
        rule=raw_measure_rule
    ),
    else_action=throw_exception
)
\end{verbatim}
}
When an error is thrown, the error message from the triggered validation rule (e.g., optionally defined as a part of \texttt{qubit\_val\_rule}, \texttt{return\_val\_rule}, or \texttt{resonator\_val\_rule}) can be included as part of the error, providing users with actionable information to modify their programs.

\subsection{Apply passes to programs}
A \texttt{PassSequence} is a construct that consists of passes in the form of:
{\small
\begin{verbatim}
pass_sequence = PassSequence(
    passes=[analysis_pass, validation_pass]
)
\end{verbatim}
}
The \texttt{pass\_sequence} defined above is a piece of data representing the validation of \texttt{raw\_measure}. When applying \texttt{pass\_sequence}, the inference engine in SHARP traverses the program AST and applies all the passes in the pass sequence sequentially to each visited node. Upon evaluating the rules in a pass sequence, the \texttt{attribute} and \texttt{kernel} arguments of the rules allow the inference engine to validate a specific constraint. The inference engine gathers information from the AST based on the \texttt{attribute} and applies checks on the attribute values based on the \texttt{kernel}. The results of the rules are then combined through the rule stacking structure into a final outcome of True or False. The outcome of the rules then triggers actions in the passes to perform the validation.

\subsection{Adding and updating the feature}
To add support for \texttt{raw\_measure} on QC1 to the quantum computing service, we add \texttt{pass\_sequence} as data to the SHARP ISA description of QC1.

Consider a case where, months later, the manufacturer of QC1 updates the hardware system and removes the resonator input variable (e.g., defaults \texttt{\$3} to use \texttt{R1} and disallows user selection). In such a case, the SHARP framework only requires that the existing rule object \texttt{raw\_measure\_rule} is replaced with the object \texttt{raw\_measure\_rule\_v2}, defined below.

{\small
\begin{verbatim}
raw_measure_rule_v2 = ALL(
    qubit_val_rule,
    return_val_rule
)
\end{verbatim}
}

\section{Discussion}
\subsection{Language decoupling}
In order to apply the kernel of a rule against an attribute of a program element, SHARP first needs to extract the attributes from the program element. This requires parsing the program and retrieving values from the properties of AST nodes. As a consequence, the overall workflow of SHARP must be dependent on the particular language in which the program is written.

\begin{figure}[htbp]
\centering
\includegraphics[width=\linewidth]{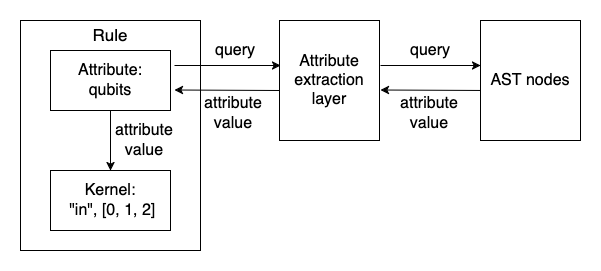}
\caption{Decoupling from the language and AST implementation.}
\label{fig:language_decoupling}
\end{figure}

In SHARP, the language dependency is decoupled by an attribute extraction layer, as illustrated in Fig.~\ref{fig:language_decoupling}. All the dependencies on the language, such as the parser and the AST implementation, are confined to the attribute extraction layer. The extraction layer has a list of members, each representing an attribute we may want to validate. Each of them is associated with an extractor. When an extractor is applied, the extracted information is sent back to the rule, which then triggers the kernel to evaluate.

For example, OpenQASM \cite{Cross2022} supports a Python implementation of the AST. The AST node for quantum gate calls is
{\small
\begin{verbatim}
@dataclass
class QuantumGate(QuantumStatement):
    modifiers: List[QuantumGateModifier]
    name: Identifier
    arguments: List[Expression]
    qubits: List[
        Union[IndexedIdentifier, Identifier]
    ]
    duration: Optional[Expression] = None
\end{verbatim}
}
OpenQASM and QIR \cite{Svore2018} may also implement parsers and ASTs in C++ which may have different names and types for the node and its attributes. Each attribute extraction layer maps attributes from a specific AST implementation to fields in SHARP based on their semantic meaning, such as mapping the \texttt{arguments} attribute in the above AST node implementation to \texttt{Instruction.inputs} in SHARP.

The part of the framework behind the extraction layer (i.e., rules, rule stacks, and passes) is language-agnostic. When targeting a different language or replacing the AST or the parser, we only need to implement a new attribute extraction layer. The rest of the framework can remain unchanged.

\subsection{Program-level constraints} \label{discussion:program_level_constraints}
In Section~\ref{sec:rule_based_static_analysis}, we showed an example of applying SHARP to support a \texttt{raw\_measure} instruction. The constraints for \texttt{raw\_measure} are at the instruction level, and the validation is based on information from a single program element. In fact, SHARP can support constraints beyond the instruction level. In order to use program-level information for validation, SHARP stores program-level results of analysis passes. In Section~\ref{subsec:Analysis_driven_validation}, the AST node data is introduced to contain information that is only accessible when visiting the node. In SHARP, there are two more data storage objects, differing in their life cycle and scope: (1) a global key-value table that is accessible by all passes and persists throughout the validation process, and (2) a pass key-value table that is only accessible to one pass and persists from the start to the end of the pass.

Below, we describe two examples of program-level constraints and explain how SHARP can support them.
\subsubsection{Instruction count}
In quantum hardware, a \texttt{measure} instruction typically takes a significantly longer time to execute compared to quantum gate instructions \cite{Karalekas2020}. If there are too many mid-circuit \texttt{measure} instructions in a program, qubits will decohere, and there will be no meaningful results. Therefore, a quantum computer can set a limit on the maximum number of \texttt{measure} instructions. Assume this limit is \texttt{M}.

This constraint can be supported by a pass sequence with two passes.
\begin{itemize}
    \item An analysis pass that checks if an AST node corresponds to a \texttt{measure} instruction. If so, it increments the value in the \texttt{measurement\_count} field of the pass key-value table by 1.
    \item A validation pass that checks if the value in the \texttt{measurement\_count} field is greater than \texttt{M}. If so, it throws an error.
\end{itemize}

\subsubsection{Order between instructions}
Some instructions may only be used when another instruction already exists. For example, an instruction \texttt{classical\_control} may only be valid if the control predicate is evaluated by a mid-circuit measurement, \texttt{mcm}. In such an example, an \texttt{mcm} instruction must precede a \texttt{classical\_control} instruction.

This constraint can be supported by a pass sequence with two passes.
\begin{itemize}
    \item An analysis pass that checks if an AST node corresponds to an \texttt{mcm} instruction with a specific predicate. If so, it sets the value in the \texttt{seen\_mcm} field of the pass key-value table to True.
    \item A validation pass that checks if an AST node corresponds to a \texttt{classical\_control} instruction with the same predicate. If it does, and if \texttt{seen\_mcm} is False, it throws an error.
\end{itemize}

\subsection{Feature access control}
A quantum computing service can support feature access control at the account level, ensuring that access to specific devices or features is granted based on user roles. For example, experimental features can be restricted to users who opt into beta programs. These users gain early access to new functionalities with the understanding that the features may be unstable, helping collect valuable feedback and enabling rapid iteration. In the framework of SHARP, because ISA descriptions are stored as standalone data, the management of feature access control is the management of data. When new access levels are added, they do not require adding logic or managing code paths in the underlying service components. SHARP enables a data-driven mechanism to regulate feature availability.

\subsection{Performance and scalability}
Quantum hardware typically supports a small set of native gates. For example, Rigetti Ankaa-3 supports three native gate types (Rx, Rz, and ISWAP), IQM Garnet supports two (PRx and CZ), and IonQ Aria-1 supports three (GPi, GPi2, and MS). Hardware constraints are generally tied to these native gates and a constant number of additional rules for control flow or device-level constraints. Thus, the number of rules grows linearly with the number of native gates, $N_g$. Given a quantum program with $N_L$ lines of code, the inference engine in SHARP applies rule stacks to each relevant AST node. The resulting validation complexity is $\mathcal{O}(N_g N_L)$, as each line may be associated with one or more rules, depending on its instruction type.

To evaluate performance in practice, we benchmarked a Python implementation of SHARP on a MacBook M1. We used a rule stack representing the ISA of Rigetti Ankaa-2 which supports four native gate types (Rx, Rz, ISWAP, CZ). Each gate is associated with 3 to 5 constraints, totaling 15 leaf rules. SHARP validated OpenQASM programs at a rate of approximately 30 microseconds per line. In comparison, parsing the same set of programs on the same compute environment took roughly 200 microseconds per line, using the default parser of OpenQASM \cite{Cross2022}. This indicates that rule-based validation introduces only a modest overhead relative to other stages.

Looking ahead, as programs grow beyond several hundred lines or rule sets become more complex, further optimization may be required. Several approaches are available to improve scalability:
\begin{enumerate}
    \item High-performance implementation: SHARP is currently implemented in Python for flexibility. Rewriting performance-critical components in a compiled language such as C++ or Rust could significantly reduce runtime.
    \item Rule stack compilation for solver-based inference: For very complex or interdependent constraints, the rule stack could be translated into logical formulas and validated using SAT or SMT solvers. This would allow SHARP to scale beyond simple pattern checks into formal constraint verification.
    \item Pipelining and parallelization: Multiple quantum programs can be validated concurrently across threads or distributed systems. Furthermore, pipelining allows stages such as program ingestion, validation, QPU execution, and result processing to overlap between different programs.
\end{enumerate}

\subsection{Applicability to fault-tolerant quantum computing}
While SHARP is motivated by near-term quantum hardware with rapidly evolving ISAs, the framework is also applicable to fault-tolerant quantum computing (FTQC) systems. FTQC platforms are expected to implement logical operations over error-corrected qubits, often within a constrained set of fault-tolerant gate primitives (e.g., Clifford+T \cite{Bravyi2005} or lattice surgery \cite{Litinski2019}). These logical-level constraints can vary depending on the underlying code (e.g., surface code, color code) or the architectural layout, and SHARP can encode such constraints as rules. Furthermore, as fault-tolerant protocols evolve to support more efficient logical operations or alternative decoding strategies, SHARP's rule-based mechanism allows validation logic to adapt without modifying the core engine. For example, enforcing distance-based scheduling rules, T-count limitations, or logical qubit layout constraints can be implemented as ISA rules, providing early validation of FTQC program suitability.

\section{Conclusion}
In this work, we introduced SHARP, a rule-based, hardware-configurable static analysis framework designed for quantum computing services. By decoupling hardware-specific validation from core static analysis functionalities, SHARP enables support for evolving quantum hardware without requiring extensive code modifications. Through a rule-based validation mechanism, the framework enforces program constraints while maintaining adaptability across different quantum ISAs.

Through its modular and analysis-driven architecture, SHARP supports instruction- and program-level validations and feature access control. The framework's design not only streamlines the onboarding of new hardware features but also promotes maintainability and extensibility for future quantum programming paradigms, including fault-tolerant systems. Additionally, the language-agnostic design of SHARP ensures compatibility across various quantum programming interfaces, thereby broadening its applicability within the quantum computing ecosystem.

\section*{Acknowledgment}
The authors thank Abe Coull, Daniela Becker, Gandhi Ramu, Jacob Feldman, Cody Wang, Ramanathan Ramanathan, Shobhit Srivastava, Vivek Dwivedi, and Yunong Shi for fruitful discussions.

\bibliographystyle{IEEEtran}
\bibliography{export}

\end{document}